\documentclass[sigconf]{acmart}

\iftrue
\copyrightyear{2022}
\setcopyright{licensedusgovmixed}
\acmConference{}
\acmPrice{}
\acmDOI{}
\acmISBN{}
\else
\copyrightyear{2022}
\acmYear{2022}
\setcopyright{licensedusgovmixed}\acmConference[EuroMPI/USA'22]{EuroMPI/USA'22: 29th European MPI Users' Group Meeting}{September 26--28, 2022}{Chattanooga, TN, USA}
\acmBooktitle{EuroMPI/USA'22: 29th European MPI Users' Group Meeting (EuroMPI/USA'22), September 26--28, 2022, Chattanooga, TN, USA}
\acmPrice{15.00}
\acmDOI{10.1145/3555819.3555820}
\acmISBN{978-1-4503-9799-5/22/09}
\fi

\usepackage{listings}
\usepackage{graphicx}

\newcommand{\hide}[1]{}
\newcommand\THREAD[1]{\texttt{MPI\_THREAD\_\-#1}}

\newif\ifdraft
\draftfalse
\ifdraft
  \newcommand{\rajeev}[1]{{\textcolor{red}{ Rajeev: #1 }}}
\else
  \newcommand{\rajeev}[1]{}
\fi

\begin{document}

\title{MPIX Stream: An Explicit Solution to Hybrid\\ MPI+X Programming}

\author{Hui Zhou}
\affiliation{\institution{Argonne National Laboratory} \country{Lemont, IL 60439, USA}}
\author{Ken Raffenetti}
\affiliation{\institution{Argonne National Laboratory} \country{Lemont, IL 60439, USA}}
\author{Yanfei Guo}
\affiliation{\institution{Argonne National Laboratory} \country{Lemont, IL 60439, USA}}
\author{Rajeev Thakur}
\affiliation{\institution{Argonne National Laboratory}\country{Lemont, IL 60439, USA}}

\begin{abstract}
    The hybrid MPI+X programming paradigm, where X refers to threads or GPUs, has gained
    prominence in the high-performance computing arena. This corresponds to a trend
    of system architectures growing more heterogeneous. 
    The current MPI standard only specifies the compatibility levels between MPI and
    threading runtimes. No MPI concept or interface exists for applications
    to pass thread context or GPU stream context to MPI implementations explicitly. 
    This lack has made performance optimization complicated in some cases and impossible
    in other cases.
    We propose a new concept in MPI, called MPIX stream, to represent the general
    serial execution context that exists in X runtimes. MPIX streams can be
    directly mapped to threads or GPU execution streams.
    Passing thread context into MPI allows implementations to precisely map the
    execution contexts to network endpoints.
    Passing GPU execution context into MPI allows implementations to directly operate
    on GPU streams, lowering the CPU/GPU synchronization cost.
\end{abstract}


\maketitle

\section{Introduction}
Modern high-performance  computing  applications are more and more dependent on additional runtimes besides
MPI to manage the increased number of cores per node and myriad limited on-node
resources such as shared memory, network interfaces, and computational accelerators.
Increasingly, applications are being deployed by using a hybrid MPI+X model,
where X refers to a threading runtime such as OpenMP or an accelerator runtime
such as CUDA.

The first step in MPI+X was to make MPI compatible with a threading runtime. Since 1997, MPI has
introduced four thread compatibility levels: \THREAD{SINGLE},
\THREAD{FUNNELED}, \THREAD{SERIALIZED}, and \THREAD{MULTIPLE}.
When the appropriate thread level is chosen, threaded applications can work
correctly with MPI without MPI specifically acknowledging the runtime.
 MPI implementations also can be made GPU aware. Recent
MPICH~\cite{mpich}, MVAPICH~\cite{mvapich}, and Open MPI~\cite{openmpi} releases are able to detect GPU buffers without hints from users and
make MPI work without a GPU-specific MPI interface. The application can
benefit from an MPI+GPU compatibility level similar to the MPI thread
levels. Currently, the GPU compatibility level is simply assumed.

While MPI+Threads is successful on the compatibility side of MPI+X, the
performance side has been a multi-decade struggle.
With MPI+Threads---in particular, at the \THREAD{MULTIPLE} thread level---applications today are still likely to meet dismal performance. This 
performance is due to the extra critical sections introduced by MPI
communications.
Much research
has been done on both the application side~\cite{wang2019multi} and implementation
side~\cite{amer2015,OMPICRI-19,Rohit-20} to address
the performance of MPI+Threads. To reach good performance, applications need to
make sure that the communications can happen concurrently, and the implementations
need to map the communications to multiple communication channels to allow the
communication to proceed in parallel. Without an explicit MPI interface,  making the latter mapping to match the application layer concurrency remains an art. 
Mismatch will result in either incorrect results or the introduction of  extra thread
contention and bad performance.

The performance story of MPI+GPUs is different from that of MPI+Threads. Accelerators
typically require special runtime to coordinate between CPU and accelerator
executions. The launching and synchronization between CPU context and
accelerator context are carried out by the accelerator runtime. A key
performance factor here is how to minimize the launching and synchronization
cost. To optimize the performance, we need MPI operations to be enqueued 
to an accelerator execution context and then let the accelerator runtime 
manage its actual execution. In order to realize this new mode of MPI
operations, new MPI interfaces that work directly with accelerator execution
context are needed.

A common theme from the pursuit of performance in MPI+X is the need for MPI to
have the concept of execution context.
In this paper we  survey the current status of MPI+Threads and MPI+GPUs
and propose a new MPI concept, called MPIX stream, that can be used to
represent execution context from other runtimes. MPIX stream  allows
explicit coordination for MPI+Threads and enables direct GPU runtime
operation for MPI communications.

\section{Background}

\subsection{MPI's Execution Model}
Among the four MPI thread levels, only  \THREAD{MULTIPLE} requires
clarification on its execution model. The other three are all serial execution
model enforced by applications.
For \THREAD{MULTIPLE}, the current guiding principle is as follows:
``When a thread is executing one of these (MPI) routines, if another
concurrently running thread also makes an MPI call, the outcome will be as if
the calls executed in some order''~\cite{mpi40}. In another words, MPI is an
outcome-dictated serial execution model.

A na\"ive implementation of the serial execution model is to impose a global
critical section for every MPI call and    yield only during its progress
loop. With the global critical section, if two threads concurrently call  MPI
communication functions, both threads  not only are serialized at the point of the MPI
operation but also incur a significant cost from the synchronization,
resulting in performance worse than if all communications are
called from a single thread.
However, an implementation is allowed to ``optimize,''  making
some parts or whole communications parallel as long as the  outcome is not affected. 

What is an outcome in MPI? Unfortunately, this is not clearly specified in the MPI standard and 
thus is a source of ambiguity and debate. Nevertheless, some consensus has been reached on what
is and what is not an outcome. For example, a message delivery
order is not an MPI outcome. A second sequentially issued message is allowed
to be delivered before the first one. On the other hand, a message matching
order is an MPI-defined outcome. Two sequentially issued sends that both
match the same receive are guaranteed to match the first one before the second
one. If we ignore outcomes that are outside the MPI standard, then an MPI
implementation can execute some of the communications in parallel as long as
there is no matching order between them. For example, messages issued from
different communicators are matched independently. If assertions that no
wildcard tag matching will be used, then messages using different tags can
also be carried out in parallel.

This strategy raises  two issues. First, the consensus is not
universal.  Certainly outcomes occur outside those specified by the MPI
standard. For example, what if the serialization or parallelization of the
communication is an application-intended outcome? Second, communicators or
tags are not perfect identification of concurrent execution context. Using
communicators or tags to sideload the expression of parallelism may result in
convoluted code yet still not necessarily be able to achieve perfect parallelization.

\subsection{Network Endpoints}
Modern high-speed interconnection fabrics are designed with capabilities to
support communication by multiple execution threads. Generally, this is done
by allowing separate fabric resources to be allocated. In this paper we
refer to these allocated fabric resources as network endpoints. Communications
can be carried out concurrently from separate endpoints.
The network endpoints are abstractions over hardware capability and  may
 include software contexts such as address table, message queues, and completion
event queues. In libfabric~\cite{website:libfabric}, a network endpoint may be
represented by a domain, an endpoint, or a scalable endpoint.
In UCX~\cite{website:ucx}, a network endpoint is typically represented by a UCP worker.

MPI implementations that utilize network endpoints also need to allocate
their own internal communication context to isolate global states that are
needed during a communication. For best performance, these
implementation-level contexts need to be matched to the network endpoints.
Both MPICH and Open MPI have implemented such communication contexts. In MPICH, it is
referred to as the virtual communication interface (VCI)~\cite{Rohit-21}. In
Open MPI, it is referred to as the communication resources instance
(CRI)~\cite{OMPICRI-19}. In this paper we generally refer to these contexts as
network endpoints.

Network endpoints are a finite resource. More endpoints beyond a hardware's
capability  will be serialized at the hardware anyway and  will  incur
more overhead in managing the multiple endpoints. A limit is often imposed
by a network library and sometimes by a network driver. It is common to have a
limit matching the number of cores in a node.

Concurrent access to a single network endpoint is not allowed, or it will
result in data race and state corruption. Thus, a critical section around the
access of each network endpoint is necessary unless it can be
guaranteed that  concurrent usages will not occur.

\subsection{Nonlocal Nature of Communication}
While both thread contexts and network endpoints are local process concepts,
a communication necessarily involves a pair of network endpoints from both the local process and the remote
process. When one does not specify a network endpoint in a communication, as is
the case with the current MPI standard, the implementation chooses a default
network endpoint for both the local process and remote process. If this
default choice is a constant choice, then all communications are serialized on
both the sender side and the receiver side. Implicit schemes or semi-explicit
schemes via hints can be used to hash the network endpoints' choice in order to achieve
parallel communications. The hashing algorithm must be deterministic and
consistent for both the sender side and receiver side. If the sender side sends a
message to a remote endpoint that does not expect to receive it, either the message
will get lost or heavy synchronization will occur in order to move the
message to the receiving context.

Implicit hashing schemes often employ certain enforcement policies. A typical policy is to
enforce a one-to-one mapping of endpoints. If we assign a sequential id to
each network endpoint in a process, then a one-to-one mapping  allows
communications only between network endpoints with the same id. With this policy,
network endpoints can be easily determined by hashing common values between
sender and receiver, such as communicator id, sender and receiver ranks, and
tags. 

Another policy is to allow the sender to send from any network endpoint but 
receive only from a default network endpoint. With this policy, the sender side can
easily achieve concurrent sends by hashing local information or even by random
assignment. Since messages are all received by a single network
endpoint, however, the overall message rates are limited by the single receiving
thread.

\begin{figure}
    \includegraphics[width=\columnwidth]{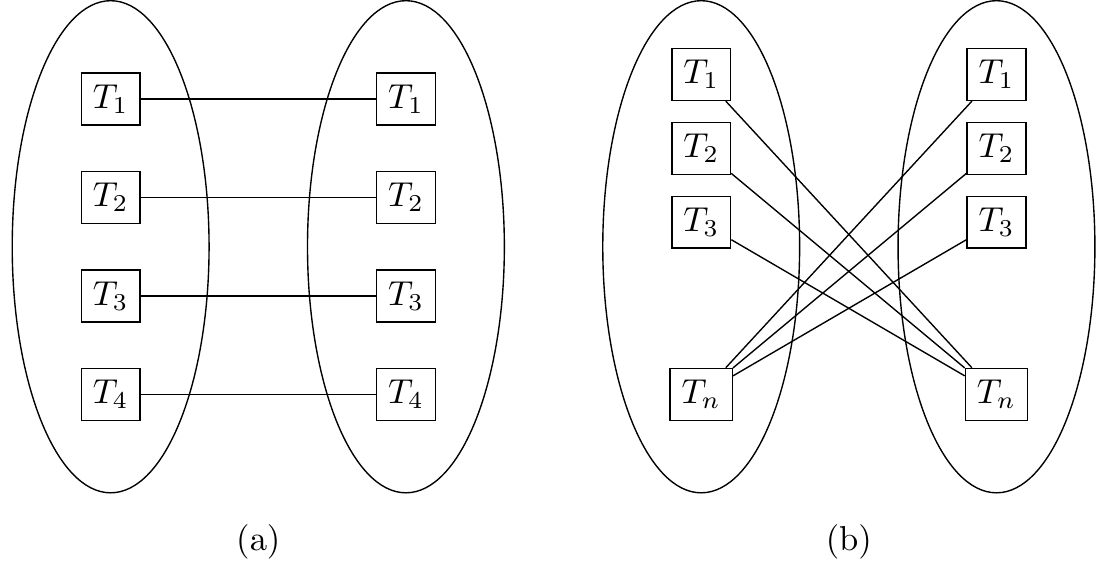}
    \caption{Typical MPI+Threads communication patterns: (a)  one-to-one
    pairwise mapping; (b)  N-to-1 mapping.}
    \label{fig:thread_mapping}
\end{figure}

\begin{figure}
    \includegraphics[width=0.6\columnwidth]{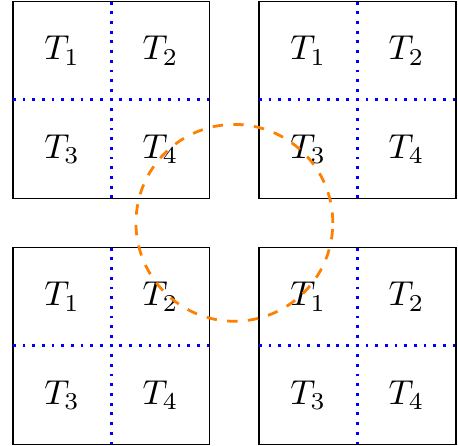}
    \caption{Communication patterns in a 2-D stencil partition.}
    \label{fig:stencil_mapping}
\end{figure}

The two policies match to the two common communication patterns
illustrated in Figure~\ref{fig:thread_mapping}: the one-to-one pattern and the
N-to-1 pattern. In a one-to-one pattern, one thread from one process 
communicates only to one thread in other processes. An example of a one-to-one
pattern is the stencil application, where a partition for a single thread only
shares a halo region with another thread in the neighbor processes (see Figure~\ref{fig:stencil_mapping}).
Note that the pairing of threads may depend on the geometry layout and may not
correspond to the ordering of the thread numbers.
In an N-to-1 pattern, multiple threads may send messages to other processes,
but a single thread in each process is dedicated to receive all messages.
An example of an N-to-1 pattern is a task-based application, where multiple
threads run tasks and generate events and where a single progress thread polls
and responds to events.

\subsection{GPU Queuing Stream}
Graphics processing units (GPUs) are  common components today in  high-performance computing clusters. GPU programming imposes its own execution
model. A common concept in GPU programming is an abstract execution queue.
In NVIDIA's CUDA runtime, this is referred to as a CUDA stream.
Unlike a thread, calls do not directly run on the execution queue. Instead,
 the operations are enqueued, and the GPU runtime will dispatch the operations to GPU
kernels asynchronously. The execution queue can be generalized as an
execution graph with explicit dependency. An execution graph may allow the GPU
runtime more flexibility in optimizations. A sequential queue, on the other
hand, often contains explicit synchronizations that may prevent optimization.

With GPU-aware MPI implementations~\cite{Wang-GPU-14}, MPI functions can be called to send from
or receive into GPU memory directly. However, MPI functions are still
implemented to execute entirely in a CPU context. Thus, full CPU/GPU
synchronizations are necessary. When the application relies on
the GPU running computations, the full CPU/GPU synchronization imposes a great performance
penalty.

To break the performance bottleneck, MPI needs to enable partial CPU/GPU
synchronization,  directly embed synchronization into the GPU kernel, or
explicitly expose dependency by using a facility such as a CUDA graph.
One way to do so is to pass the GPU runtime execution context to MPI
so that the MPI implementation can directly operate under the GPU context. In
the case of CUDA, this may mean  passing the CUDA stream object into MPI send
and receive functions. 

In the GPU execution context, examples including CUDA stream and SYCL queue are often
asynchronous. This poses another challenge on  working with MPI's execution
model. How do we assess matching order when the issuing order of the operation
itself is not deterministic? Unlike in the MPI+Threads case where a default
serialization model makes sense, MPI+GPUs will need a new concept to accommodate
the foreign execution context.

\section{Proposal of MPIX Stream}
We propose a new MPI object, called MPIX stream, and a set of new APIs to
allow users to explicitly identify their thread context and GPU queuing
context. We use the prefix MPIX instead of MPI to refer to objects and functions that are proposed in this paper and not (yet) officially part of MPI.

\subsection{MPIX Stream}
First, we introduce MPIX stream as an abstract concept to facilitate a direct mapping from
user-level runtime execution contexts to MPI execution contexts.
To MPI, an MPIX stream represents a local serial execution context.
Any runtime execution contexts outside MPI, as long as the serial semantic is strictly followed, can be
associated to an MPIX stream. Examples include kernal threads, user-level threads,
GPU queuing streams, or even code across multiple threads with serialized synchronizations.

To illustrate, in the following two pseudocode listings, we use \texttt{MPIR\_SEND\_ON\_STREAM}
to represent an arbitrary MPI operation, and \texttt{stream\_1} represents a specific MPIX stream object
that is associated with the MPI operation. Because in both listings, the MPI operations are strictly
serialized, both are valid usages of MPIX stream.

\begin{lstlisting}[caption={Pseudocode using MPIX stream within a single
thread}, label={lst:usestream}]
{
    /* within a single thread */
    MPIR_SEND_ON_STREAM(stream_1, msg1);
    MPIR_SEND_ON_STREAM(stream_1, msg2);
}
\end{lstlisting}
\begin{lstlisting}[caption=Pseudocode using MPIX stream from two threads with explicit thread synchronization]
{
    /* Thread 1 */
    MPIR_SEND_ON_STREAM(stream_1, msg1);
    THREAD_BARRIER();
}
{
    /* Thread 2 */
    THREAD_BARRIER();
    MPIR_SEND_ON_STREAM(stream_1, msg2);
}
\end{lstlisting}
Note that pseudocode is used in these examples, and in particular, \texttt{MPIR\_SEND\_ON\_STREAM}
is not a proposed API. We will describe the actual APIs later in this section.
The examples also illustrate that it is the users' responsibility to map their
thread context to the MPIX stream.
\rajeev{It is not clear how the examples highlight that. It may be a requirement, but not obvious just from the examples.}
This relieves the burden of MPI having to synchronize
between threads, which can be thread runtime dependent.

To create an MPIX stream, one calls \texttt{MPIX\_Stream\_create}.
\begin{lstlisting}[language=C]
int MPIX_Stream_create(MPI_Info info, MPIX_Stream *stream)
\end{lstlisting}
Info hints can be used to create implementation-supported special streams, for
example, a CUDA stream. Otherwise, \texttt{MPI\_INFO\_NULL} can be used. Whether the
returned stream is backed by  distinct network endpoints is implementation
dependent. For MPI thread contexts, we recommend  allocating unique
network endpoints for each new stream. This approach allows programming the application  with
a predictable performance. The  implementation should return failure if it runs out
of network endpoints. With unique endpoints and the strict serial execution
context, the implementation may safely skip critical sections in the communication
path. At the extreme end of the strong scaling, even an uncontended critical
section can be too expensive.

The implementation may also assign a single network endpoint to multiple MPIX
streams, allowing applications to create more streams than available network
endpoints. For example, network endpoints can be assigned to a newly created
stream in a round-robin fashion. This may provide flexibility for some
applications. We note that a per-endpoint critical section is necessary to prevent
concurrent access to network endpoints.

Because network endpoints are finite resources, users should free the
stream to make the resource available for future allocation.
\begin{lstlisting}[language=C]
int MPIX_Stream_free(MPIX_Stream *stream)
\end{lstlisting}
The network resource  can be deallocated only when all the operations using the
stream have been completed. We imagine 
that whether a
deallocation is successful is important  feedback for users.  In particular, a
failed or delayed deallocation may prevent a future \texttt{MPIX\_Stream\_create} from
succeeding. Thus, \texttt{MPIX\_Stream\_free} may fail with an appropriate error code if the
internal resource deallocation cannot be completed.


GPU streams emphasize  lightweight synchronization between the CPU and
accelerator. Thus, having concurrent CPU communications may not be as
important as in multithreaded programming. In addition, an implementation may
choose to use a dedicated CPU thread to progress all GPU stream
communications. Thus, GPU streams are likely to be assigned with duplicate
network endpoints. On the other hand, having
dedicated network endpoints for GPU stream-related communications is likely beneficial so that the
progress thread  needs to poll only a few network endpoints rather than polling
global progress and contending with traffic on other CPU threads.

For backward compatibility, \texttt{MPIX\_STREAM\_NULL} is defined. Operations on
\texttt{MPIX\_STREAM\_NULL} have the same semantics as do conventional operations without
explicit streams.

\subsection{Passing Opaque Binary Info Hints}
Currently, an \texttt{MPI\_Info} object  supports values only as strings. A GPU queuing object
not only is not a string but is often opaque to the user. For example, is a CUDA
stream an integer or a pointer, or could it be neither? To pass an opaque
binary as a string, we need some encoding scheme that users can use to encode and
implementations can consistently decode. We propose a new MPI function for this purpose.

\begin{lstlisting}[language=C]
int MPIX_Info_set_hex(MPI_Info info, const char *key, void *value, int vallen)
\end{lstlisting}

An implementation can choose any binary to ASCII encoding to implement this
function. For completeness, there should be a corresponding
\texttt{MPIX\_Info\_get\_hex} function. For this proposal, however, we focus on how
to let users pass the GPU queuing object into MPI.  Since getting the object back from MPI is not critical, we are not proposing the retrieving
function here.

\subsection{MPIX Stream Communicator}
Once an MPIX stream is created, it is possible to define MPI operation APIs that
directly use the stream as an argument to each operation. This has two drawbacks, however.
First, we will need to create a new API for every MPI operation, from
\texttt{MPI\_Send}, \texttt{MPI\_Isend}, to MPI collectives and MPI remote memory access (RMA) operations. Not only does it
take considerable effort to maintain an inflated standard, but it is also a burden for
users to learn these APIs. Second, it is not sufficient to add only a local
stream to an operations argument. We must add an argument for a remote
stream as well unless we want to restrict to an arbitrary policy. For collectives, this
may require an array of stream arguments, one for every participating process. Even for
one-sided RMA operations, a specific targeting stream may be critical if the user
wants to dedicate a progress thread to drive passive progress. Unlike local
streams, identifying remote streams per operation is  cumbersome for the  user.

Thus we propose the MPIX stream communicator.
\begin{lstlisting}[language=C]
int MPIX_Stream_comm_create(MPI_Comm parent_comm, MPIX_Stream stream, MPI_Comm *stream_comm)
\end{lstlisting}

This is a collective operation. Stream information from all processes or its
network endpoint address can be Allgathered and stored locally. All
conventional MPI operations can be issued to a stream communicator without
additional parameter changes.

If the \texttt{parent\_comm} is also a stream communicator, it is treated as a normal
communicator. That is, the stream attached to the \texttt{parent\_comm} is discarded
in the new communicator.

Each stream is still local to each process, and streams do not need to agree in
any aspects between processes. In particular, any process is allowed to use
\texttt{MPIX\_STREAM\_NULL} in constructing the stream communicator. The operation
context is dictated by each process's attached stream. Unless it is
\texttt{MPIX\_STREAM\_NULL}, a strict serial context per stream is assumed.

One also can mix normal streams with GPU streams. Once again, the
operation context is dictated by its local stream. It is possible to have a
collective where one process is posting immediate operations when another
process is enqueuing its corresponding operation, and it is up to the GPU runtime to
asynchronously execute them.

\subsection{MPIX Enqueue APIs}
Because an attached local stream dictates the operation mode, it is feasible to use
the same conventional MPI operations without syntax change for GPU enqueue
functions. 
However, there are subtle semantic differences between an enqueued operation and
a non-enqueued operation.  To an
advocate of explicit coding style, having, for example,  \texttt{MPI\_Send} enqueue to a GPU
stream can seem a very bad code. Thus we propose the following specific enqueue
APIs.
\rajeev{What do these enqueue functions actually do? When does the send or recv happen?}

\begin{lstlisting}[language=C]
int MPIX_Send_enqueue(const void *buf, int count, MPI_Datatype datatype, int dest, int tag, MPI_Comm comm)
int MPIX_Recv_enqueue(void *buf, int count, MPI_Datatype datatype, int source, int tag, MPI_Comm comm, MPI_Status *status)
int MPIX_Isend_enqueue(const void *buf, int count, MPI_Datatype datatype, int dest, int tag, MPI_Comm comm, MPI_Request *request)
int MPIX_Irecv_enqueue(void *buf, int count, MPI_Datatype datatype, int source, int tag, MPI_Comm comm, MPI_Request *request)
int MPIX_Wait_enqueue(MPI_Request *request, MPI_Status *status)          
int MPIX_Waitall_enqueue(int count, MPI_Request array_of_requests[], MPI_Status array_of_statuses[])
\end{lstlisting}

These routines have identical signatures as their conventional counterparts.
It is an error to call the enqueue functions if the communicator is not a
stream communicator or does not have a local GPU stream attached. We
intentionally omitted \texttt{MPIX\_Waitsome} and \texttt{MPIX\_Waitany}, as well as the various
test functions, because the nondeterministic nature of these functions does
not work with the enqueue environment.
Moreover, \texttt{MPIX\_Waitall\_enqueue} must have requests all issued on
the same local stream.

There are semantic differences between enqueuing APIs and conventional APIs. For example,
\texttt{MPIX\_Send\_enqueue}, as with all enqueuing APIs, returns immediately after registering the operation. A separate progress
thread, which may be the GPU runtime thread, will initiate and complete the communication
asynchronously. This is different from the conventional nonblocking API, e.g., \texttt{MPI\_Isend}. An implementation of \texttt{MPI\_Isend}
may process the message buffer and initiate the communication immediately before return, while
both \texttt{MPIX\_Send\_enqueue} or \texttt{MPIX\_Isend\_enqueue} will only process the buffer and initiate communication
after the previous enqueued operations complete.
Both \texttt{MPIX\_Send\_enqueue} and \linebreak \texttt{MPIX\_Isend\_enqueue} will return immediately after registering the operation, but
\texttt{MPIX\_Isend\_enqueue} allows the following enqueued functions to proceed before the communication
completes with a corresponding \texttt{MPIX\_Wait\_enqueue}.
It may appear confusing since both the traditional non-blocking APIs and the new enqueue APIs are asynchronous. It is worth noting that they are dealing with two orthogonal kinds of synchronicity. Traditional non-blocking APIs are mostly concerned with synchronizing data in the message buffer, while the enqueue APIs are concerned with synchronization between execution contexts. The former is synchronized by calls such as \texttt{MPI\_Wait}, while the latter is sychronized by calls such as \texttt{cudaStreamSynchronize}. In particular, with the addition of the enqueue APIs, GPU synchronization calls, such as \texttt{cudaStreamSynchronize}, are no longer needed for message data or communication synchronizations.

The enqueue APIs can be extended to collectives and RMA functions. All the extended
enqueue functions will have identical function signatures as their conventional counterparts.
They are not listed here due to the large number of them. For collectives, if some of the processes
are not associated with an enqueuing stream, then those processes should call the conventional
non-enqueue API. For RMA, enqueuing operations include window synchronizations.

\subsection{MPIX Multiplex Stream Communicator}
The stream communicator works well for codes that  communicate only to a
single thread on a remote process. The communicator effectively constructs a
thread communication group and, similar to legacy MPI code, can achieve
concurrent multithread performance without any change to the code.

On the other hand, when a thread needs to communicate with multiple threads
of a remote process, it may be necessary to create and manage
multiple stream communicators. Doing so can be cumbersome and become unmanageable
 quickly. For example, two processes each with 4 threads will need 16
stream communicators to achieve an all-to-all communications.

To address this usability issue, we propose the MPIX multiplex stream
communicator.
\begin{lstlisting}[language=C]
int MPIX_Stream_comm_create_multiple(MPI_Comm parent_comm, int count, MPIX_Stream local_streams[], MPI_Comm *stream_comm)
\end{lstlisting}

With a multiplex stream communicator, each process can attach multiple local
streams.
To use the MPIX multiplex stream communicator, we propose the following point-to-point APIs.
\begin{lstlisting}[language=C]
int MPIX_Stream_send(const void *buf, int count, MPI_Datatype datatype, int dest, int tag, MPI_Comm comm, int src_idx, int dst_idx)
int MPIX_Stream_recv(void *buf, int count, MPI_Datatype datatype, int source, int tag, MPI_Comm comm, int src_idx, int dst_idx, MPI_Status *status)
int MPIX_Stream_isend(const void *buf, int count, MPI_Datatype datatype,    int dest, int tag, MPI_Comm comm, int src_idx, int dst_idx, MPI_Request *request)
int MPIX_Stream_irecv(void *buf, int count, MPI_Datatype datatype, int source, int tag, MPI_Comm comm, int src_idx, int dst_idx, MPI_Request *request)
\end{lstlisting}

These APIs allow users to explicitly address local and remote streams via an
index. This index can be thought of as a rank for threads.
Unlike the source rank in send and destination rank in receive, both \texttt{src\_idx} and \texttt{dst\_idx} are needed in the argument list since the communicator here does not uniquely identify the local thread.
\texttt{MPIX\_ANY\_INDEX} can be used to support a wildcard receive.

The multiplex stream communicator is especially useful for an N-to-1
communication pattern, where one or a few polling threads are responsible for
receiving messages sent from any other thread.  Without multiplex stream
communicators, one must create multiple single-stream communicators
and have the polling thread poll each communicator in turn. With multiplex
stream communicators, the polling thread  needs to poll only a single communicator.

\section{Comparison with Previous Efforts}
Researchers have been addressing  the paradigm of MPI+Threads for two decades. 
Many previous efforts have been reported,  and this proposal draws experience from these past
efforts. In this section we compare our work with the previous proposals and alternative
solutions.

\subsection{Implicit Method}
The implicit method~\cite{OMPICRI-19,Rohit-20}
builds on the thesis that MPI  already has a sufficient mechanism
 to allow users to express the inherent parallelism in their
communications. Based on the outcome-dictated serial execution model, any
operations that do not affect MPI-specified outcomes are candidates for deserialization.
In particular, users can express parallelism using separate communicators. 

Using distinct communicators for different thread communication groups is
 similar to the usage of stream communicators in our proposal. However,
to write code to fully take advantage of implicit methods requires knowledge of 
the particular MPI implementation, and the behavior is never guaranteed. For
example, for a one-to-one pattern, if the implementation uses a different
policy other than per-communicator mapping, the performance will not be
ideal. In contrast, with our proposal, the mapping to network endpoints is
explicitly identified with  the MPIX stream, and its behavior is guaranteed.

When the communication does not fit into a one-to-one pattern, one thread may
need to use more than one communicator. In a typical implementation of the
implicit method, where network endpoints are assigned to communicators in a
round-robin fashion, when each thread uses multiple communicators, two threads
may still be assigned with the same network endpoint despite using different
communicators. Using  MPIX stream, however, network endpoint assignments are explicit.
Thus, using multiple communicators per thread will not be an issue.

One must use fine-grained critical sections with the implicit method.
When there is no thread contention, there will be a slight performance penalty
compared with using a global critical section because of the overhead of using more
locks. On the other hand, correct usage of MPIX stream allows implementation to
skip locking altogether, resulting in a performance gain.

\subsection{MPI Endpoints Proposal}
The MPIX stream proposal has its roots in the MPI endpoints proposal~\cite{Endpoints-14}. The
multiplex stream communicator is nearly identical to an endpoints
communicator. Both proposals allows direct addressing of individual endpoint
or thread context.

A key flaw in the endpoint proposal, from our view, is the inflation of
thread context into virtual processes.
To a multithread programmer, a process and a thread are separate concepts.
Between threads the memory is shared, and thus there is no need for explicit data
exchange. Between processes, explicit messaging is needed, and that is where
MPI is used. Thus, the process concept and its identification using ranks are
important. With the endpoints proposal, the process become less identifiable.
Users may have to manage their own endpoint ranks to process the rank translation
table. The endpoints proposal also makes interthread messages equally accessible
as interprocess messages. Since users rarely need interthread messages, this
inflation makes MPI more difficult to understand and use.

In contrast, a multiplex stream communicator maintains
the address via process ranks plus the thread index. This fits the model of
multithreaded application naturally, and thus it is easier to learn and use.

\subsection{MPI-4 Partitioned Communication}
Partitioned communication is a new addition to the MPI-4 standard. One of its
motivations is to provide simpler and more effective multithread
optimization~\cite{Finepoints-19}. Partitioned communication has an explicit
init stage where implementations can set up strategy and decide network
endpoints mapping to partitions. The actual communications can be triggered by
\texttt{MPI\_Pready} calls, which can occur concurrently or out of order.
\rajeev{Some missing words in the second half of previous sentence}

This should be compared with sending multiple messages from multiple threads,
each message corresponding to a single partitioned data. Using explicit streams,
users always can control the thread-to-message mapping, thus achieving the desired
parallelism. Partitioned communication, on the other hand, is still an
implicit mapping mechanism, although through the init stage implementations
may be able to achieve better mapping than implicit static mapping can.

However, partitioned communication focuses on the optimization of a single
parallel region with a single coordinated functionality where individual
partitioned data is part of bigger data that can be described with a single
message. It does not solve the concurrency issue when there are other messages
besides the partitioned message or when the message cannot be equally
partitioned. For example, an application may need to exchange data on an irregular region, which cannot use a
partitioned scheme.

 MPIX stream, on the other hand, lets users explicitly control thread mapping and thus 
orchestrate communications across multiple areas and even
orchestrate beyond single parallel regions. The latter is critical when
applications have dedicated threads taking care of some of the communication
needs.

\subsection{MPI-4 Sessions}
Another new addition to the MPI-4 standard is MPI Sessions~\cite{session2019}. Intuitively from the
name, an MPI session is intended as a local context to encapsulate all MPI
local objects. It can support independent, including concurrent, MPI operations
from different MPI sessions.

Most general-purpose objects can be specialized for a particular purpose. During
proposal development, the MPI Forum quickly discovered that MPI sessions can be
used to solve all kinds of issues. Simply by attaching special attributes or
hints to a session, it can become specific enough to 
address almost any issue that requires specific context. 

For example, an MPI session can be used as an MPIX stream. To do so, we need to create an
MPI session with \THREAD{SERIALIZED} thread level, then identify the
session with a local thread context. We can also attach a GPU queueing object
to the session using info hints. The session can proceed to create
communicators, which essentially are stream communicators. From here, all the
semantics proposed for stream communicators can proceed. The only missing
functionality is the multiple stream communicator.

Although this solution via MPI Sessions works, it is not intuitive and is
convoluted. It is difficult to connect the name ``session'' as an equivalent to a
thread, stream, or network endpoint. To educate users on such session usages
will be difficult. To use  sessions in this way, each session needs to go through a
bootstrapping stage until arriving at a communicator. Compared with the pattern
that we initialize,  setting up the common part including the initial collective
communications, and then creating specialized communicators for special code
regions, the code using an MPI session can be difficult to manage because the code is
less separated by purposes.

Unless the MPI standard specifies  special usage, an
implementation is unlikely to optimize the usage of an MPI session as an MPIX stream.

\subsection{NVIDIA Collective Communication Library (NCCL)}
NCCL~\cite{NCCL} is a library providing selected point-to-point and collective communications
between GPU devices. NCCL's design is optimized for dense multi-GPU systems, while
MPI is geared toward interprocess communications across many nodes in a cluster.
NCCL uses a stream-based enqueue-only API, which our proposed enqueue APIs are directly
modeled after. NCCL being a CUDA-specific library can directly use CUDA stream in
its interface. This tie to a specific external runtime is undesirable for MPI, which
need be usable for a variety of GPU runtimes. With \texttt{MPIX\_Stream}, the specific
tie is limited to the info hints during stream creation, while the rest of
the APIs are portable across different GPU runtimes.

Beside the enqueuing streams, NCCL focuses on communications between GPU devices, and
its communicator is formed by a collection of GPU devices. On the other hand, MPI communications
are interprocess by default, although self messages are also allowed. An MPI communicator
is always formed by a group of processes, and each rank addresses a single process.

Lastly, NCCL, being a specialized library, only supports contiguous buffers with intrinsic
datatypes, and its operations are limited to a selected set of operations. Our proposed
enqueue APIs, on the other hand, work for MPI datatypes, and can be readily extended
to all MPI operations.

\subsection{Alternative Proposal for GPU Enqueues}
Alternative proposals to add GPU enqueue operations to MPI are to directly add
the GPU queue objects to point-to-point operations as extra arguments.

\begin{lstlisting}[language=C]
int MPIX_Send_enqueue(const void *buf, int count, MPI_Datatype datatype, int dest, int tag, MPI_Comm comm, enum MPIX_QUEUE_TYPE type, void *stream)
int MPIX_Recv_enqueue(void *buf, int count, MPI_Datatype datatype, int source, int tag, MPI_Comm comm, enum MPIX_QUEUE_TYPE type, void *stream, MPI_Status *status)
int MPIX_Isend_enqueue(const void *buf, int count, MPI_Datatype datatype, int dest, int tag, MPI_Comm comm, enum MPIX_QUEUE_TYPE type, void *stream, MPI_Request *request)
int MPIX_Irecv_enqueue(void *buf, int count, MPI_Datatype datatype, int source, int tag, MPI_Comm comm, enum MPIX_QUEUE_TYPE type, void *stream, MPI_Request *request)
int MPIX_Wait_enqueue(MPI_Request *request, enum MPIX_QUEUE_TYPE type, void *stream, MPI_Status *status)
int MPIX_Waitall_enqueue(int count, MPI_Request array_of_requests[], enum MPIX_QUEUE_TYPE   type, void *stream, MPI_Status array_of_statuses[])  
\end{lstlisting}

This is essentially the same as skipping the MPIX stream creation and stream
communicator creation and moving the info hint from the stream directly to each enqueue
operation. If we  look at only  the listed functions, this is a simpler and more
direct way of achieving it.

The MPIX stream uses separate steps to create the stream and then the stream
communicators. Thus it has more opportunities for implementations to validate
and optimize. It is also easily extensible by accepting more info hints. It 
readily extends the functionality to collectives and one-sided communications
without extra APIs. Also, the stream creation and communicator construction
provide error check opportunities so users can choose to use fallback algorithms
if necessary.
Cosmetically, passing an opaque object via a void pointer reference is less
desirable.

Namashivayam et al.\ in a recent study~\cite{hpe-triggerop} proposed a new data object, \texttt{MPIX\_Queue}, which is similar to \texttt{MPIX\_Stream}
but limited as an abstraction over GPU stream objects. They proposed a similar set of APIs using \texttt{MPIX\_Queue} as a direct argument. Compared
to our proposal, it shares the benefit of \texttt{MPIX\_Stream} abstraction, but lacks the extensibility from the stream communicator construction.

\section{Prototype Implementation}
We have implemented a prototype of the stream APIs proposed in this paper. The
prototype is available  in the MPICH 4.1a1 release.

\subsection{Mapping VCI to MPIX Stream}
MPICH internally already maintains a pool of virtual communication interfaces. With the per-VCI critical section model, each VCI uses separate mutexes and
accesses dedicated network endpoints. Communications from separate VCIs can be
fully concurrent. For a detailed discussion of VCIs, see \cite{Rohit-20}.

MPICH currently supports implicit VCI hashing for \THREAD{MULTIPLE} using
traditional APIs.
For our prototype implementation, we separate the pool of VCIs into an implicit
pool and an explicit pool. The size of each pool can be controlled by the user via
MPI tool interface control variables. The total number of VCIs is
limited by both available network endpoints and software limits. More VCIs
will demand larger tables in various internal data structures, as well as
heavier address exchange during initialization, so we advise users to set the
size of VCI pools according to their application usage. For example, if the
application is not using the stream APIs proposed in this paper, users should
leave the reserved VCI pool size at 0 and set the implicit VCI pool size to
match the number of threads or number of cores they are using. On the other
hand, if the application is using the stream APIs, we expect users want to
 control their stream mapping explicitly throughout the application.
Thus, they should leave the implicit VCI pool size at the default, 1, and set
the reserved VCI pool size according to the total number of allocated streams.

All the functions proposed in this paper are implemented; however, not all
functionality is complete. In particular, one-sided operations are not
explicitly stream-aware. A window created by using a stream communicator will
behave like a conventional communicator with implicit VCI assignment. Point-to-point functions and collective functions, including nonblocking and
persistent variations, are fully stream-aware.

\subsection{GPU Enqueue APIs}
The GPU enqueue functionality is  implemented only for CUDA and only through
the explicit point-to-point enqueue functions. The work to extend the
functionality to collectives and one-sided communication is ongoing.

The current implementation uses CUDA's \texttt{cudaLaunchHostFunc} to enqueue the MPI
operation to the CUDA stream. Not all GPU runtimes provide host enqueue functions;
and even with CUDA, this is not optimal. The current CUDA implementation
incurs a heavy switching cost for \texttt{cudaLaunchHostFunc}.

A better implementation may use a dedicated host thread to the progress operation
queue and  enqueue only the event triggers or event synchronizations to the
kernel queues. Having MPI  launch separate progress threads is another area
worth exploring. Having each runtime spawn hidden progress threads behind
the user's knowledge is often not optimal.

\subsection{Results}
\begin{figure}
    \includegraphics[width=\columnwidth]{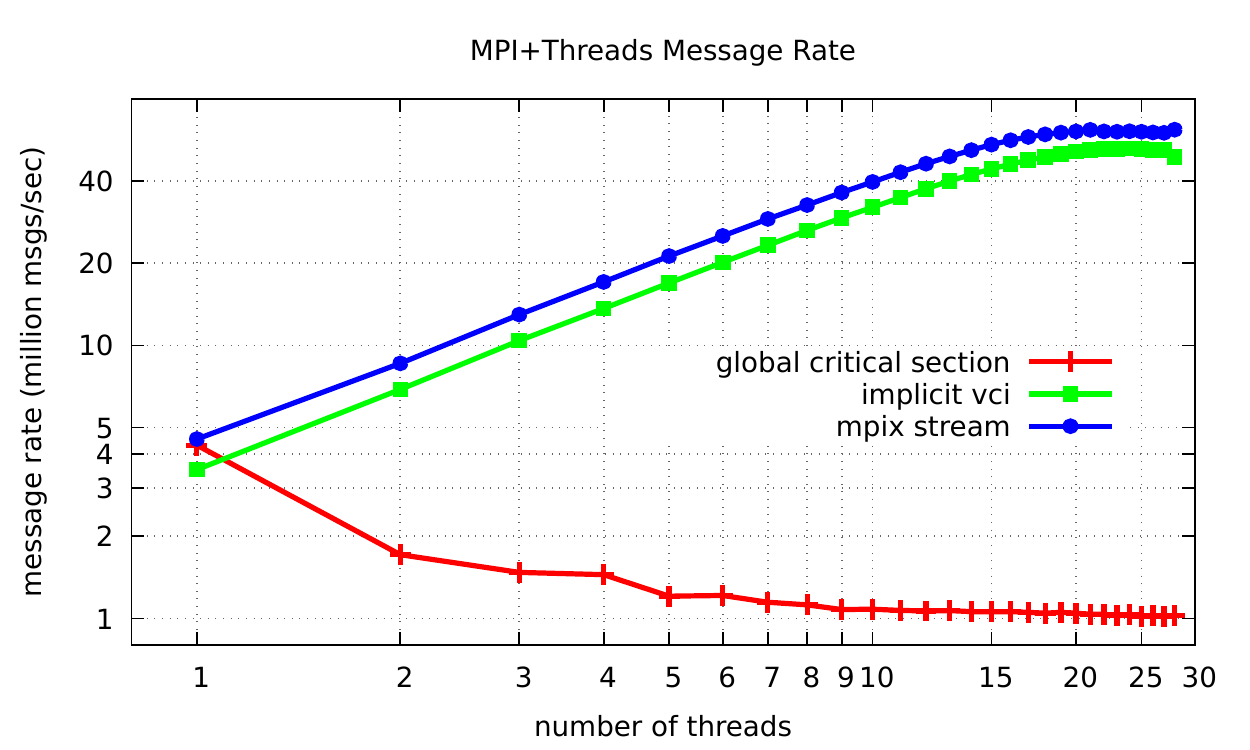}
    \caption{Multithread message rate on 8-byte messages using
    MPI\_Isend/MPI\_Irecv.The message rate using MPIX\_Stream is around $20\%$ higher than with implicit VCIs.}
    \label{fig:msgrate}
\end{figure}

To verify our implementation, we conducted a microbenchmark measurement on an
Intel Skylake cluster in the Joint Laboratory for System Evaluation (JLSE) at Argonne
National Laboratory. The nodes in the cluster are connected by Mellanox InfiniBand 
EDR. The microbenchmark launches a number of threads, and each thread then sends 8-byte
messages to a corresponding thread on another process. Each thread
communicates using a per-thread communicator. The results are shown in
Figure~\ref{fig:msgrate}.
Three measurements were taken. The first was with MPICH configured to use the global
critical section. Shown in the red curve, we see that the total message rate drops as
soon as more threads  start to compete for the critical section.

The green curve shows the message rate with MPICH configured to use the per-VCI critical
section. Here traditional MPI communicators are used, and thus MPICH is implicitly
hashing the communications using different VCIs. The microbenchmark is designed
to achieve perfect implicit hashing, and we see good scaling as we increase
the number of threads. Note that the message rate with a single thread is actually
smaller than the corresponding message rate with the global critical section. The reason is that the per-VCI critical sections are finer grained and it often takes multiple
critical sections along the communication path---in particular, the receive
path and progress 
engine---for each message to complete. Even without
contention, the extra locking and unlocking hurt the performance.

The third curve, shown in blue, is from rewriting the benchmark to use the
proposed stream communicators. Each stream communicator is attached with a unique
MPIX stream object per thread. Because the semantics of MPIX stream guarantees
a serial execution context, our implementation is able to completely remove
locking, resulting in around $20\%$ gain in the total message rate up to 20 threads.

In our current implementation, atomic variables and atomic operations are still
used to reference count request objects and completion flags. Even uncontended
atomics hurt performance in these microbenchmarks. Unfortunately, the current
MPICH code structure made it difficult to switch off the atomic operations.
This issue remains on our  to-do list.

\section{Examples}
In this section we show two examples to illustrate how the proposed APIs will be used.
In Listing \ref{stream_single} we show an example hybrid MPI+OpenMP program for a one-to-one thread communication pattern.
Each thread is represented by a unique \texttt{MPIX\_Stream} and uses a dedicated stream communicator to send and receive messages.
A good implementation can ensure these communications happen concurrently without incurring extra locking cost.

\begin{lstlisting}[language=C, label=stream_single, caption=Example using MPIX stream communicator
with OpenMP]
#define NT 4

int main(void) {
    int rank;
    int tl;
    MPI_Init_thread(NULL, NULL, MPI_THREAD_MULTIPLE, &tl);
    MPI_Comm_rank(MPI_COMM_WORLD, &rank);

    MPIX_Stream streams[NT];
    MPI_Comm comms[NT];
    for (int i = 0; i < NT; i++) {
        MPIX_Stream_create(MPI_INFO_NULL, &streams[i]);
        MPIX_Stream_comm_create(MPI_COMM_WORLD, streams[i], &comms[i]);
    }

    #pragma omp parallel num_threads(NT)
    {
        int id = omp_get_thread_num();
        char buf[100];
        int tag = 0;
        if (rank == 0) {
            MPI_Send(buf, 100, MPI_CHAR, 1, tag, comms[id]);
        } else if (rank == 1) {
            MPI_Recv(buf, 100, MPI_CHAR, 0, tag, comms[id], MPI_STATUS_IGNORE);
        }
    }
    for (int i = 0; i < NT; i++) {
        MPIX_comm_free(&comms[i]);
        MPIX_Stream_free(&streams[i]);
    }

    MPI_Finalize();
    return 0;
}
\end{lstlisting}

In Listing \ref{gpu_enqueue} we show an example MPI+CUDA program using CUDA's asynchronous stream enqueue facility. It is a simple vector computation, SAXPY. Process 0 generates a portion of the data and sends it to process 1, which launches the kernel to do the computation after receiving the data. All memory copies, MPI send/receive, and computation kernels are asynchronously launched to a user-supplied CUDA stream.

\begin{lstlisting}[language=C, label=gpu_enqueue, caption=Example using MPIX
stream for CUDA stream enqueue operations]
const float a_val = 2.0;
const float x_val = 1.0;
const float y_val = 2.0;

__global__
void saxpy(int n, float a, float *x, float *y)
{
  int i = blockIdx.x*blockDim.x + threadIdx.x;
  if (i < n) y[i] = a_val * x[i] + y[i];
}

int main(void)
{
    cudaStream_t stream;
    cudaStreamCreate(&stream);

    int rank;
    MPI_Init(NULL, NULL);
    MPI_Comm_rank(MPI_COMM_WORLD, &rank);

    float *x, *y, *d_x, *d_y;

    MPI_Info info;
    MPI_Info_create(&info);
    MPI_Info_set(info, "type", "cudaStream_t");
    MPIX_Info_set_hex(info, "value", &stream, sizeof(stream));

    MPIX_Stream mpi_stream;
    MPIX_Stream_create(info, &mpi_stream);

    MPI_Info_free(&info);

    MPI_Comm stream_comm;
    MPIX_Stream_comm_create(MPI_COMM_WORLD, mpi_stream, &stream_comm);

    /* Rank 0 sends x data to Rank 1, Rank 1 performs a * x + y and checks result */
    if (rank == 0) {
        x = (float*)malloc(N*sizeof(float));
        for (int i = 0; i < N; i++) {
            x[i] = x_val;
        }
        MPIX_Send_enqueue(x, N, MPI_FLOAT, 1, 0, stream_comm);

        free(x);
    } else if (rank == 1) {
        y = (float*)malloc(N*sizeof(float));
        cudaMalloc(&d_x, N*sizeof(float));
        cudaMalloc(&d_y, N*sizeof(float));

        for (int i = 0; i < N; i++) {
            y[i] = y_val;
        }
        cudaMemcpyAsync(d_y, y, N*sizeof(float), cudaMemcpyHostToDevice, stream);
        MPIX_Recv_enqueue(d_x, N, MPI_FLOAT, 0, 0, stream_comm, MPI_STATUS_IGNORE);
        saxpy<<<(N+255)/256, 256, 0, stream>>>(N, a, d_x, d_y);

        cudaMemcpyAsync(y, d_y, N*sizeof(float), cudaMemcpyDeviceToHost, stream);

        cudaFree(d_x);
        cudaFree(d_y);
        free(y);
    }
    
    MPI_Comm_free(&stream_comm);
    MPIX_Stream_free(&mpi_stream);

    cudaStreamDestroy(stream);
    MPI_Finalize();

    return 0;
}
\end{lstlisting}

\section{Summary}
We have surveyed the current status of MPI+Threads and MPI+GPUs. Both will need
an explicit interface in MPI to allow better arrangements between non-MPI
runtimes and MPI. We proposed a new MPI concept, called MPIX stream,
and a set of new APIs that allow users to communicate their external execution
context to MPI implementations in a general and reliable way. The proposed
APIs are implemented  in the MPICH 4.1a1 release. Example codes for typical
application patterns are provided for reference.

\begin{acks}
Special thanks to Jim Dinan from NVIDIA for insightful discussions and his
effort in leading the MPI Forum hybrid working group. We thank MPI developer
teams from Intel corporation and Hewlett Packard Enterprise for feedback and
the members of the MPI Forum for past efforts and on-going discussions related
to this work. We gratefully acknowledge the computing resources provided and operated by the Joint Laboratory for System Evaluation (JLSE) at Argonne National Laboratory. This research was supported by the Exascale Computing Project (17-SC-20-SC), a collaborative effort of the U.S.\ Department of Energy Office of Science and the National Nuclear Security Administration, and by the U.S. Department of Energy, Office of Science, under Contract DE-AC02-06CH11357. 
\end{acks}

\bibliographystyle{ACM-Reference-Format}
\bibliography{references}

 \end{document}